\documentclass[a4paper,12pt]{article}
\usepackage{amssymb,amsmath}
\usepackage{bm}%
\usepackage{stmaryrd}
\usepackage[pdftex]{hyperref}
\hypersetup{
	colorlinks=true,
	citecolor=blue,
	linkcolor=blue,
	urlcolor=orange,
}
\usepackage{cite}
\usepackage{comment}
\usepackage{xcolor}

\def\Z2{\mathbb{Z}_2^2}%

\begin{document}

\title{Integration on minimal $\mathbb{Z}_2^2$-superspace and emergence of space}

\author{N. Aizawa \thanks{Corresponding author: aizawa@omu.ac.jp} and  Ren Ito
	\\[10pt]
	Department of Physics, Graduate School of Science, \\ Osaka Metropolitan University, \\
	Nakamozu Campus, Sakai, Osaka 599-8531, Japan}

\maketitle
\thispagestyle{empty}

\vfill
\begin{abstract}
	We investigate the possibilities of integration on the minimal $\mathbb{Z}_2^2$-superspace. Two definitions are taken from the works by Poncin and Schouten and we examine their generalizations. It is shown that these definitions impose some restrictions on the integrable functions. We then introduce a new definition of integral, which is inspired by our previous work, and show that the definition does not impose restrictions on the integrable functions. An interesting feature of this definition is the emergence of a spatial coordinate which means that the integral is defined on $\mathbb{R}^2$ despite the fact that the $(0,0)$ part of the minimal $\mathbb{Z}_2^2$-superspace is $\mathbb{R}. $ 
\end{abstract}

\clearpage
\setcounter{page}{1}

\section{Introduction}

The aim of the present work is to investigate integration on $\Z2$-superspace which  is the simplest example of higher graded extensions of supermanifolds. 
The local coordinates of a supermanifold are given by real numbers and Grassmann numbers which form a commutative superalgebra ($\mathbb{Z}_2$-graded algebra). One may replace the $\mathbb{Z}_2$-grading structure of supermanifolds by $ \mathbb{Z}_2^n := \mathbb{Z}_2 \times  \mathbb{Z}_2 \times \cdots \times \mathbb{Z}_2 $ ($n$ times) and the resulting object is called  $\mathbb{Z}_2^n$-manifolds. 
The coordinates of a $\Z2$-superspace are elements of an abelian $\Z2$-superalgebra \cite{RW1,RW2} (see also \cite{Ree,sch1}) so that  the $\Z2$-superspace is a natural extension of superspace used widely in supersymmetric theories. 
Geometry of supermanifolds has been understood very well (see e.g. \cite{Var,PonSch}), while higher graded supergeometry is currently the subject of active research \cite{PonSch,CGP1,CGP2,CGP3,CGP4,CoKPo,Pz2nint,BruIbar,BruPon,BruIbarPonc,BruGraRiemann,BruGrabow,CoKwPon,BruIbarPon2,BruGrabow2,MohSal}. 
There are many open questions in higher graded supergeometry. It might be surprising that integration on $\mathbb{Z}_2^n$-manifolds is one of the open questions.  
The main obstacle to integration is the existence of an exotic coordinate which is not nilpotent but anticommute with fermionic coordinates.  
Such coordinate exists for any $\mathbb{Z}_2^n$-manifolds for $ n > 1$, thus how to define integration on $\Z2$-superspace is a highly nontrivial problem.
Two inequivalent definitions have been proposed so far \cite{Pz2nint,PonSch}, however, there is no generally accepted definition in the community of mathematicians.

In this paper, we consider the minimal $\Z2$-superspace consisting of four coordinates, i.e., one real, one exotic and two nilpotent. 
Our main results are summarized as follows: (i) we examine the possible definitions along the line of \cite{Pz2nint,PonSch} and show that the well-definedness imposes some constraints on integrand in addition to the component functions of the integrand are compactly supported. 
Recall that the compact support is also the integrability condition on  supermanifolds. 
(ii) we propose a new definition of integral which does not impose any additional constraints on integrands. 
This definition is inspired by our previous work \cite{AIKT} and the integral is defined in two dimensional Euclidean space, despite the fact that the $(0,0)$ part of the minimal  $\Z2$-superspace is $ \mathbb{R}. $  
The emergence of a new coordinate stems from that one may regard the square of the exotic coordinate as a real number. 

Our study of integration is motivated by recent extensive studies of $\Z2$-supersymmetric theories \cite{Bruce,BruDup,AAD,AAd2,AKTcl,AKTqu,DoiAi1,DoiAi2,brusigma,bruSG,Topp,Topp2,AIKT}. 
The $\Z2$-supersymmetry opens up a new area of research of great interest. 
For instance, $\Z2$-supersymmetric quantum mechanics introduces a new type of parafermions and parabosons and it has detectable differences from the ordinary supersymmetric quantum mechanics \cite{Topp,Topp2}.  We point out that para-particles have been simulated recently by using a trapped ion \cite{ParaP}. Thus, it would be important to study para-particles from various perspectives. 
The $\Z2$-extension of the sine-Gordon equation was introduced and the integrablity of it has been shown  \cite{bruSG}. 
This suggests the existence of a new class of integrable systems characterized by $\Z2$-supersymmetry. 
We also mention that a classification of $\Z2$-superdivision algebras has been made  \cite{FraZhan}. 
It is an interesting problem how the new superdivision algebras relate to other mathematics or nature just as we have observed in the 10-fold way.

Similar to the standard supersymmetry, $\Z2$-superspace and $\Z2$-superfields will be very useful tools for analyzing $\Z2$-supersymmetry. To establish this formalism,  integration on $\Z2$-superspace is a necessary element. 
Of course, integration on $\mathbb{Z}_2^n$-manifold is essential for higher graded geometry and the present work for the minimal $\Z2$-superspace is the first step towards it. 

The paper is organized as follows. We recall the basics of $\Z2$-superspace and the calculus on it in \S \ref{SEC:Pre}, focusing on the foundations of integration theory. 
In \S \ref{SEC:Integral}, we first give an explicit formula of the $\Z2$-Berezinian. 
Then, three different definitions of integration on the minimal $\Z2$-superspace are examined. 
Two of them are generalizations of those introduced in \cite{Pz2nint,PonSch} and it is shown in \S \ref{SubSEC1} and \S \ref{SubSEC2} that those definition requires some restriction on integrable functions. On the contrary, the definition we introduce in \S \ref{SubSEC3} does not impose any restrictions on integrable functions. 
What is peculiar about the definition is that it is accompanied by an emergence of space. 
The last section present a summary of the results, outlining the further line of investigation. 
\section{Minimal $\Z2$-superspace and $\Z2$-graded calculus} \label{SEC:Pre}

We begin with the definition of $\Z2$-\textit{commutative} algebra. 
Each element of the algebra $ X_{\vec{a}}$ has \textit{degree} which is  an element of $ \vec{a} \in \Z2 := \{  \ (0,0), \ (0,1), \ (1,0), \ (1,1)\ \}$. The product of two element $X_{\vec{a}} $ and $ Y_{\vec{b}} $ obey the following rule
\begin{equation}
	X_{\vec{a}} Y_{\vec{b}} = (-1)^{\vec{a}\cdot\vec{b}} Y_{\vec{b}} X_{\vec{a}}
\end{equation}
where  $ \vec{a} \cdot \vec{b}$ be the standard inner product of two dimensional vectors and the degree of the product $ X_{\vec{a}} Y_{\vec{b}} $ is given by $ \vec{a}+\vec{b}.$ 

The $\Z2$-commutative algebra is turned into an abelian $\Z2$-superalgebra by introducing a $\Z2$-graded Lie bracket
\begin{equation}
	\llbracket X_{\vec{a}}, Y_{\vec{b}} \rrbracket := X_{\vec{a}} Y_{\vec{b}} - (-1)^{\vec{a} \cdot \vec{b}} Y_{\vec{b}} X_{\vec{a}}.
\end{equation}
One may readily see that the $\Z2$-graded Lie bracket is nothing but commutator (anticommutator) for $ {\vec{a}} \cdot {\vec{b}}$ even (odd). 

We consider the \textit{minimal} $\Z2$-superspace throughout this paper and the coordinates of it are denoted by
\begin{equation}
	x \; (0,0), \qquad \xi \; (0,1), \qquad \eta \; (1,0), \qquad z \; (1,1). \label{Coordinates}
\end{equation}
The coordinates are elements of a $\Z2$-commutative algebra according to the indicated degree. 
More explicitly, all the following (anti)commutators vanish:
\begin{equation}
	[x, \bullet\;], \quad \ \{\xi, \xi\}, \quad [\xi,\eta], \quad \{\eta, \eta\}, \quad \{\xi,z\}, \quad \{\eta, z\}.
\end{equation}
One may regard $x$ as a real number. $\xi $ and $\eta $ are nilpotent, which allow us to treat them as commuting fermions. 
$z$ is an exotic bosonic coordinate in the sense that it is not nilpotent and anticommutes with fermionic variables. 
The main issue of the present work is how to define an integral over this exotic coordinate. 

A function $F$ on the minimal $\Z2$-superspace is a formal power series of the exotic coordinate $z$:
\begin{equation}
	F(x,z,\xi,\eta) = \sum_{k=0}^{\infty} \sum_{\alpha,\beta=0,1} g_{k\alpha\beta}(x) z^k \xi^{\alpha} \eta^{\beta}.
\end{equation}
A component function $ g_{k\alpha\beta}(x) $ is a  function on $\mathbb{R}$ and has nontrivial degree. When $\deg(F) = (0,0)$, $ \deg(g_{k\alpha\beta}) = (k+\beta,k+\alpha) \mod 2.$ 
We assume throughout the paper that all the component functions are $C^{\infty}(U)$ and \textit{compactly supported} in a domain $ U \subseteq \mathbb{R}.$  

We also introduce derivatives with respect to the coordinates \eqref{Coordinates}
\begin{equation}
	\partial_x := \frac{\partial}{\partial x} \; (0,0), \qquad 
	\partial_{\xi} \; (0,1), \qquad \partial_{\eta} \; (1,0), \qquad \partial_z \; (1,1)
\end{equation}
which has the degree same as the corresponding coordinate. 
They are also elements of a $\Z2$-commutative algebra and they form a $\Z2$-superalgebra together with the coordinates:
\begin{equation}
	\llbracket \partial_{X_{\vec{a}}},  \partial_{X_{\vec{b}}} \rrbracket = 0, 
	\qquad
	\llbracket \partial_{X_{\vec{a}}},  X_{\vec{b}} \rrbracket = \delta_{\vec{a},\vec{b}}.
\end{equation}

We employ the formulation of integration over the $\Z2$-superspace developed in \cite{Pz2nint,PonSch}. 
The volume form is given by
\begin{equation}
	dx dz \otimes \partial_{\xi} \partial_{\eta} 
\end{equation}
which is legitimised  by a cohomological argument \cite{Covolo}. 
The volume form implies that integration over the fermionic variables is identical to the Grassmann integral, i.e., integral and derivative are the same. 
If we change the coordinates 
\begin{equation}
	(x,z,\xi,\eta) \ \to \ (u,v,\zeta,\theta)
\end{equation}
the volume form transforms according to
\begin{equation}
	du dv \otimes \partial_{\zeta} \partial_{\theta} = dx dz \otimes \partial_{\xi} \partial_{\eta} \cdot \Gamma(J)
\end{equation}
where $\Gamma(J)$ is a $\Z2$-Berezinian of the (modified) Jacobian matrix $J$
\begin{equation}
	J := \begin{pmatrix}
	     	A & B \\ C & D 
	     \end{pmatrix}
	     :=
	    \left(\begin{array}{cc|cc}
		  \partial_x u & \partial_x v & \partial_x \zeta & \partial_x \theta
		  \\
		  \partial_z u & \partial_z v & \partial_z \zeta & \partial_z \theta
		  \\ \hline
		  \partial_{\xi} u & \partial_{\xi} v & \partial_{\xi} \zeta & \partial_{\xi} \theta
		  \\
		  \partial_{\theta} u & \partial_{\theta} v & \partial_{\theta} \zeta & \partial_{\theta} \theta
		\end{array}
		\right)
\end{equation}
defined by \cite{CGP3}
\begin{equation}
	\Gamma(J) = \frac{\det(A-BD^{-1}C)}{\det D}. \label{BerDef}
\end{equation}
The $\Z2$-Berezinian is multiplicative and $\deg(\Gamma(J)) = (0,0). $  
The expression \eqref{BerDef} looks the same as the superdeterminant, which is due to the simplicity of the minimal $\Z2$-superspace. 
In more general $\Z2$-superspace, determinant in \eqref{BerDef} is replaced with $\Z2$-determinant \cite{CGP3}. 

Let  $\beta$ be a Berezinian section defined by 
\begin{equation}
	\beta := (dx dz \otimes \partial_{\xi} \partial_{\eta}) F(x,z,\xi,\eta).
\end{equation}
The integral of $\beta$ is given by
\begin{equation}
	\int \beta := \int dx \int dz\, \partial_{\xi} \partial_{\eta} F(x,z,\xi,\eta)
	= \int dx \int dz\ \sum_{k=0}^{\infty} g_{k11}(x) z^k \label{Int0}
\end{equation}
and how to define the integral on the right-hand side is the problem discussed in this paper.
The integral has to be well-defined, which means that its value must not change when the coordinates are changed.

\section{Possible integrations on minimal $\Z2$-superspace} \label{SEC:Integral}
\setcounter{equation}{0}

To verify that the integral is well-defined, it is useful to have an explicit formula of the $\Z2$-Berezinian. The most general coordinate transformation is given by \cite{Pz2nint}:
\begin{align}
	u &= \sum_{r=0}^{\infty} \Big( f_r^u(x) z^{2r} + g_r^u(x) z^{2r+1} \xi \eta \Big),
	\nonumber \\
	v &= \sum_{r=0}^{\infty} \Big( f_r^v(x) z^{2r+1} + g_r^v(x) z^{2r} \xi \eta  \Big),
	\nonumber \\
	\zeta &= \sum_{r=0}^{\infty} \Big( f_r^{\zeta}(x) z^{2r} \xi + g_r^{\zeta}(x) z^{2r+1}\eta 	 \Big),
	 \nonumber \\
	 \theta &= \sum_{r=0}^{\infty} \Big( f_r^{\theta}(x) z^{2r} \eta + g_r^{\theta}(x) z^{2r+1} \xi \Big) 
	 \label{CoordinateChange}
\end{align}
where all the functions of $x$ such as $ f_r^u(x), g_r^u(x)$ are real and degree $(0,0). $ 
It is convenient to introduce the degree $(0,0)$ variable $ y:= z^2$ and write \eqref{CoordinateChange} in the following form:
\begin{align}
	u &= f^u(x,y) + g^u(x,y) z \xi \eta,
	\nonumber \\
	v &= f^v(x,y) z + g^v(x,y) \xi \eta,
	\nonumber \\
	\zeta &= f^{\zeta}(x,y) \xi + g^{\zeta}(x,y) z \eta,
	\nonumber \\
	\theta &= f^{\theta}(x,y) \eta + g^{\theta}(x,y) z \xi  \label{CoordinateChange2}
\end{align}
where the functions of two variables $ x, y$ are defined by 
\begin{equation}
	f^u(x,y) := \sum_{r=0}^{\infty}  f_r^u(x) y^r, 
	\qquad 
	g^u(x,y) := \sum_{r=0}^{\infty} g_r^u(x) y^r, \quad \mathrm{etc.} 
\end{equation}
and all of them are degree $(0,0). $ 
With these two variable functions, the Berezinian for \eqref{CoordinateChange2} is given by
\begin{align}
	\Gamma(J) &= \frac{J^B}{\det D} + G(x,y)  z \xi \eta,
	\nonumber \\
	\det D &= f^{\zeta}(x,y) f^{\theta}(x,y) - y g^{\zeta}(x,y) g^{\theta}(x,y),
	\nonumber \\
	 G(x,y)&= \partial_x \frac{(f^v + 2y f^v_{y}) g^u - 2f^u_y g^v}{\det D} 
	  + \partial_y \frac{2(f^u_x g^v - y f^v_x g^u)}{\det D} 
\end{align}
where the subscript $x,y$ of a function denotes the partial derivative, e.g.,  $ f^v_y := \partial_y f^v(x,y) $  and $J^B$ is the Jacobian for the  change of the bosonic coordinates
\begin{equation}
	J^B := \det 
	\begin{pmatrix}
		\partial_x u & \partial_z u 
		\\
		\partial_x v & \partial_z v
	\end{pmatrix}_{\xi=\eta=0}
    = f^u_x f^v + 2y(f^u_x f^v_y - f^u_y f^v_x).
\end{equation}

Using this formula, we examine three definitions of integration  to see what happens when we make a coordinate change.

\subsection{Integration according to \cite{Pz2nint}} \label{SubSEC1}

The Berezinian section in the transformed coordinates $(u,v,\zeta,\theta)$ is given by
\begin{equation}
	\beta = (du dv \otimes \partial_{\zeta} \partial_{\theta}) F(u,v,\zeta,\theta) 
	= (du dv \otimes \partial_{\zeta} \partial_{\theta}) \sum_{k=0}^{\infty} g_{k\alpha\beta}(u) v^k \zeta^{\alpha} \theta^{\beta}
	\label{Bsection}
\end{equation}
and we consider the following definition of integral 
\begin{equation}
	\int \beta = \int_U g_{\ell 11}(u) du \ \in \mathbb{R} \label{DefInt1}
\end{equation}
where $\ell $ is a non-negative integer. 
The $ \ell = 0 $ case has been discussed and it was shown that the well-defindness requires that $ g_{100} = 0 $  \cite{Pz2nint}. This means that $ z, $ which is a typical quantity in $\Z2$-superspace, does not contribute the integral at all. We thus also examine the case $ \ell > 0. $ 

In \eqref{DefInt1}, there is an inconsistency in the degree of the left and right sides for even $ \ell$ as $ \deg(\text{LHS}) = (0,0)$, while $ \deg(\text{RHS}) = (\ell+1,\ell+1) \mod 2.$  
In addition, the well-definedness requires some component functions to vanish,  similar to $\ell =0.$ For example,  $ g_{000} = g_{200} = 0 $ for $ \ell = 1 $ and $g_{100} = g_{300} = g_{011} = 0 $ for $ \ell =2. $ The number of vanishing component functions increases as $\ell $ becomes large. 
The vanishing component functions for general $\ell$ are given as follows:
\begin{alignat}{2}
	\ell &= 2m \ ;  &  
	\begin{cases}
		g_{2k+1\, 00} = 0, & 0 \leq k \leq m
		\\[5pt]
		g_{2k\, 11} = 0, & 0 \leq k \leq m-1
	\end{cases}
	\nonumber \\[5pt]
	\ell &= 2m+1 \ ; \ &
	\begin{cases}
		g_{2k\, 00} = 0, & 0 \leq k \leq m+1
		\\[5pt]
		g_{2k+1\, 11} = 0, & 0 \leq k \leq m-1
	\end{cases} 
    \label{list1}
\end{alignat}

To see this, we make the coordinate change \eqref{CoordinateChange2}. 
Due to the nilpotecy of $\xi, \eta$ we have the relations
\begin{align}
	g_{k\alpha\beta}(u) &= g_{k\alpha\beta}(f^u) + \frac{d g_{k\alpha\beta}}{df^u} g^u z \xi \eta,
	\nonumber \\
	v^k &= (f^v z)^k + k (f^v z)^{k-1} g^v \xi\eta,
	\nonumber \\
	\zeta \theta &= \det D \cdot \xi \eta.
\end{align}
Using these, it is straightforward to verify 
\begin{align}
	\int \beta &= \int dx dz\, T(x,z) = \int_U dx \left. \frac{1}{\ell !} \frac{\partial^{\ell} T(x,z)}{\partial z^{\ell}} \right|_{z=0},
	\label{Int1} \\
	T(x,z) &:= \sum_{k=0}^{\infty} \left[ J^B (f^v)^k g_{k11} z^k +  k\frac{J^B}{\det D} \,g^v (f^v)^{k-1} g_{k00} z^{k-1}  \right.
	\nonumber \\
	&+ \left. \left( \frac{J^B}{\det D} g^u \frac{dg_{k00}}{df^u} + G(x,y) g_{k00} \right)(f^v)^k z^{k+1} \right] 
	\label{Tfunction}
\end{align}
where the last equality of \eqref{Int1} is due to the definition \eqref{DefInt1}. 
Note that $ g_{k\alpha\beta} $ is the function of $x $ and $ y $ through $ g_{k\alpha\beta}= g_{k\alpha\beta}(f^u(x,y)). $ 

It may not be difficult to see that only the component functions in \eqref{list1}, except $g_{\ell 11}$, appear in the integral \eqref{Int1} and their derivatives are up to the first order, 
since higher order derivative of $ g_{k\alpha\beta}$ keeps the factor $z^m \; (m > 0)$ and we set $z=0.$ We give some explicit expressions for lower values of $\ell$ as an example: For $ \ell = 0$
\begin{equation}
 \int \beta = \int_U dx \left[ f^u_x f^v g_{011} + \frac{f^u_x f^v}{f^{\zeta} f^{\theta}} g_{100} \right]_{y=0}
\end{equation}
and for $ \ell = 1 $
\begin{equation}
	\int \beta = \int_U dx \left[ f^u_x (f^v)^2 g_{111} + G(x,y) g_{000} + \frac{f^u_x f^v}{f^{\zeta} f^{\theta}} \left( 2f^v g^v g_{200} + g^u \frac{dg_{000}}{df^u} \right) \right]_{y=0} 
\end{equation}
and for $ \ell = 2 $
\begin{align}
	\int \beta = \int_U dx &\left[ f^u_x (f^v)^3 g_{211} + G(x,y) f^v g_{100} + \frac{f^u_x (f^v)^2}{f^{\zeta} f^{\theta}} \left( 3 f^v g^v g_{300} + g^u \frac{dg_{100}}{df^u} \right) \right. 
	\nonumber \\
	&+ \left.   
	    \partial_y \left( f^u_x f^v \Big( \frac{g^v}{f^{\zeta} f^{\theta}}  g_{100} +   g_{011} \Big) \right)
	\right]_{y=0}.
\end{align}
In all examples, the first term on the RHS gives the correct integral and all other terms must disappear for the integral to be well-defined. The functions multiplied by $ g_{k\alpha\beta} $ are determined by the coordinate transformation \eqref{CoordinateChange2} so that they take almost arbitrary form. Therefore, the component functions $ g_{k\alpha\beta}$ itself has to vanish identically.  

The existence of the constraints on integrable functions is problematic when we construct $\Z2$-supersymmetric theory by using the superfield formalism. Superfield is a function on a $\Z2$-superspace and we integrate a function, which is constructed from superfields, on the $\Z2$-superspace to obtain a $\Z2$-supersymmetric action. 
The constraints on the integrand impose additional relations for physical fields and the relations would  be physically unacceptable in many cases.

%
\subsection{Integration according to \cite{PonSch}} \label{SubSEC2}

Let us examine another possible definition of integration on the minimal $\Z2$-superspace introduced in \cite{PonSch}. 
This definition involves changing the nature of the integrable function. We consider a function which is defined by a Laurent series
\begin{equation}
	F(u,v,\zeta,\theta) 
	=  \sum_{k=-M}^{\infty} g_{k\alpha\beta}(u) v^k \zeta^{\alpha} \theta^{\beta}
\end{equation}
where the lower bound $M \in \mathbb{N}$ is finite, but we do not fix its value. 
Then we define the integral by 
\begin{equation}
	\int \beta = \int_U g_{-\ell 11}(u) du \ \in \mathbb{R} \label{DefInt2}
\end{equation}
where $ \ell \in \mathbb{N}. $ The proposal of \cite{PonSch} is to take $ \ell = 1, $ which is inspired obviously by the residue theorem in complex analysis. 
However, we do not restrict ourselves to $ \ell = 1$ and also examine other values of $\ell,$ though there is an  inconsistency of degree similar to the definition \eqref{DefInt1}.   
Note that the definition \eqref{DefInt2} implies that $ \int \beta = 0 $ if $ \ell > M. $ 

The well-definedness imposes additional constraints on the component functions except $ M = \ell = 1.$ The existence of constraints is readily seen from a simple example. Let us consider the transformation 
\begin{equation}
	u =x, \quad v = (1+y) z + \xi \eta, \quad \zeta = \xi, \quad \theta = \eta.
\end{equation}
In this case, we have
\begin{equation}
	\det D = 1, \quad J^B = 1+3y, \quad G(x,y) = 0.
\end{equation}
It follows that
\begin{align}
	\int \beta &= \int_U dx \sum_{r=0} (-1)^r 
	\left[
	  \begin{pmatrix}
	  	\ell-1+3r \\ \ell-1+2r
	  \end{pmatrix} 
     \big( g_{-\ell-2r\,11}(x) \theta^M(\ell+2r) - \ell_1\, g_{-\ell_100}(x) \theta^M(\ell_1)\big)
	\right.
	\nonumber \\
	& \left. + 3 \begin{pmatrix}
		\ell+1+3r \\ \ell+1+2r
	\end{pmatrix}
    \big( g_{-\ell_211}(x) \theta^M(\ell_2) -\ell_3 \,g_{-\ell_300}(x) \theta^M(\ell_3) \big) \right] 
    \label{Int2}
\end{align}
where
\begin{equation}
	\ell_1 := \ell-1+2r, \quad \ell_2 := \ell+2+2r, \quad \ell_3 := \ell+1+2r 
\end{equation}
and 
\begin{equation}
	 \theta^M(x) := 
	 \begin{cases}
	 	1, & x \leq M \\[7pt] 0, & \text{otherwise}.
	 \end{cases}
\end{equation}
The first term with $r=0 $ of \eqref{Int2} gives the correct integral and all other terms must disappear which impose $ g_{k11} = g_{k00} = 0 $ for some $k$'s. 

The $ M=\ell=1$ case is exceptional, there is no additional constraints. 
To see this, note that \eqref{Tfunction} holds true for negative $k$ so that we have for $ M=1$
\begin{align}
	\int \beta &= \int dx dz\, T(x,z), 
    \\
	T(x,z) &:= \sum_{k=-1}^{\infty} \left[ J^B (f^v)^k g_{k11} z^k +  k\frac{J^B}{\det D} \,g^v (f^v)^{k-1} g_{k00} z^{k-1}  \right.
	\nonumber \\
	&+ \left. \left( \frac{J^B}{\det D} g^u \frac{dg_{k00}}{df^u} + G(x,y) g_{k00} \right)(f^v)^k z^{k+1} \right]. 
\end{align}
The $z^{-1}$ pole arises only from $ (f^v z)^{-1} $ so that
\begin{equation}
	\int \beta = \int_U dx \left[ \frac{J^B}{f^v} g_{-111}\right]_{y=0} = \int_U dx \big[ f^u_x g_{-111} \big]_{y=0}
\end{equation}
which is the correct integral after the coordinate transformation. 
However, the requirement of $ M=1$ is a strong restriction of integrable functions. 

%
\subsection{Emergence of space in integration} \label{SubSEC3}

We introduce a new definition of integral which is inspired by our previous work \cite{AIKT}.  
We return to the function defined by a Taylor series \eqref{Bsection} and write it in the following form
\begin{align}
	F(u,v,\zeta, \theta) &= \varphi_{00}(u,w) + \tilde{\varphi}_{11}(u,w)v 
	+ \big( \psi_{01}(u,w) + \tilde{\psi}_{10}(u,w) v \big) \zeta
	\nonumber \\
	&+ \big( \psi_{10}(u,w) + \tilde{\psi}_{01}(u,w) v \big) \theta
	+ \big( A_{11}(u,w) + A_{00}(u,w)v \big) \zeta \theta \label{8components}
\end{align}
where $ w:= v^2 $ is a variable of degree $ (0,0)$ which commutes with any other coordinates and the new component functions are defined by
\begin{equation}
	\varphi_{00}(u,w) := \sum_{k=0}^{\infty} g_{2k\, 00}(u) w^k, \qquad 
	\tilde{\varphi}_{11} := \sum_{k=0}^{\infty} g_{2k+1\, 00}(u) w^k, \quad \text{etc.}
\end{equation}
The suffix of the coordinate functions indicates their degree. In terms of the new component functions, the integral \eqref{Int0} is computed as
	\begin{align*}
		\int du\,dv \partial_{\zeta} \partial_{\theta} F &= \int du\,dv \big( A_{11}(u,w) + A_{00}(u,w)v \big) 
		\\
		&
		\stackrel{w=v^2}{=}  \int du \,dw \frac{1}{2}\big( A_{11}(u,w)v^{-1} + A_{00}(u,w) \big). 
	\end{align*}

Now we define the value of this integral by
\begin{equation}
	\int \beta = \frac{1}{2} \int_D du\, dw A_{00}(u,w) \ \in \mathbb{R} \label{Int3}
\end{equation}
We dropped $A_{11}(u,v)$ in this definition. As is seen from the proof below, if we keep  $A_{11}(u,v)$ then the integral is not well-defined. 

The RHS of \eqref{Int3} is defined on a domain $D \in \mathbb{R}^2, $ namely, the variable $w$ is mapped to a real number. 
The same mapping has been introduced in \cite{AIKT} through the realization of $\Z2$-commutative coordinates by the quaternionic matrices. 
Here, we observe that the same mapping, without the realization but from the nature of $w$, i.e., $\deg(w) = (0,0) $ and $[w, \bullet] = 0,$ is applicable to define the integration. 
The domain $D$ will be fixed by assuming that all the component functions, $ \varphi_{00}, \tilde{\varphi}_{11}, $ etc. are compactly supported in $D.$ Recall that the assumption is necessary for fermionic integral.  
Furthermore, one may verify that the integral \eqref{Int3} is well-defined under this assumption. To see this, we make the coordinate transformation \eqref{CoordinateChange2} and the induced transformation of $w$:
\begin{equation}
	w = (f^v z + g^v \xi\eta)^2 = y(f^v)^2 + 2 f^v g^v z \xi \eta. 
\end{equation}
It follows from the definition \eqref{Int3} that
\begin{align}
	\int \beta &= \frac{1}{2} \int_D dx\, dy \Big( J^B f^v A_{00}(x,y) + \partial_x j^x + \partial_y j^y \Big)= \frac{1}{2} \int_D dx\, dy  J^B f^v A_{00}(x,y),
	\label{Int4} \\
	j^x &= \big[ (f^v+2yf^v_y) g^u - 2f^u_y g^v \big] \frac{\varphi_{00}}{\det D},
	\nonumber \\
	j^y &= 2( f^u_x g^v - y f^v_x g^u ) \frac{\varphi_{00}}{\det D}.
\end{align}
The last equality in \eqref{Int4} is due to the assumption of compact support. 
Recalling that $J^B $ is the Jacobians for the bosonic coordinate transformations $ (x,z) \to (u,v) $ and noting that $f^v$ stems from the change $ z \to v$ in the last term of \eqref{8components},  we see that the RHS of \eqref{Int4} gives the correct integral. 

The emergence of a new spatial coordinate through integration is a curious phenomenon, but surprisingly this definition does not impose any additional constraints on integrand. 
A similar emergence of space has also been observed when the quaternionic realization is employed \cite{AIKT}. 
This realization produces an interesting two-dimensional supersymmetric model from the model on the minmal $\Z2$-superspace, and a closed string is derived if the emergent coordinate is compactified on a circle $S^1.$


\section{Conclusions}

We examined three different integration on the minimal $\Z2$-superspace. 
It was shown that two of them require integrable conditions in addition to the compact support of the component functions, while the remaining one does not. 
Furthermore, the last one \eqref{Int3} is accompanied by the emergence of a new space coordinate. The mapping of $w$ to a real number is based on naive observation that $\deg(w)=(0,0)$ and commutativity. A more mathematically rigorous formulation is desirable.  
The definition \eqref{Int3} has a great advantage when applied to physical problems, since it does not impose any additional relations on physical fields. Thus, it may be applicable to not only superfield formulation of $\Z2$-supersymmetry, but also $\Z2$-superconformal mechanics using nonlinear realizations. 

We restricted ourselves to the minimal $\Z2$-superspace in this paper. 
If we repeat the same analysis for $\Z2$-superspaces beyond the minimal one, similar integrability conditions are likely to exist for the integration of the types \eqref{DefInt1} and \eqref{DefInt2},  but not to exist for the type \eqref{Int3}. 
Moreover, one may consider another type of $\Z2$-superspaces. Recall that the $\Z2$-superspace considered in this work is an abelian $\Z2$-superalgebra and there exists  different type of $\Z2$-graded extension of Lie algebras which is sometimes called ``$\Z2$-graded Lie algebra" \cite{RW1,RW2}. Abelian $\Z2$-graded Lie algebras may define another type of $\Z2$-superspace. 
Detailed analysis on these non-commutative space has not been done yet. 
Finally, we mention that a classification of minimal $\Z2$-graded Lie (super)algebras has been made and the existence of many varieties of them has been shown \cite{KTclassification}. This leads various interesting considerations of higher graded symmetries in mathematics and physics.


\section*{Acknowledgments} 

N. A. is supported by JSPS KAKENHI Grant Number JP23K03217. 

%
%

\end{document}